\documentclass[twocolumn,prl,groupedaddress]{revtex4-2}
\usepackage{siunitx}

\usepackage{overpic}
\usepackage{float}
\usepackage{hyperref}
\usepackage{framed, color}
\usepackage{bbm}
\usepackage[dvipsnames]{xcolor}

\usepackage{amssymb}
\usepackage{amsmath}
\usepackage[normalem]{ulem}

\newcommand{\+}{\dagger}

\newcommand{\ZP}{{\textstyle\frac{1}{2}}}

\newcommand*\VLin{V_L^{\text{in}}}
\newcommand*\VRin{V_R^{\text{in}}}
\newcommand*\VLout{V_L^{\text{out}}}
\newcommand*\VRout{V_R^{\text{out}}}

\begin{document}

%\preprint{}

%\title{Maybe add theory in the high impedance title so people do not think we realised it}
 \title{Transmon in a semi-infinite high-impedance transmission line \textemdash \  appearance of cavity modes and Rabi oscillations }

\author{ E. Wiegand$^{1}$ , B. Rousseaux$^{2}$ ,  G. Johansson$^{1}$ }
\email[]{wiegand@chalmers.se}
\affiliation{$^{1}$ Applied Quantum Physics Laboratory, Department of Microtechnology and Nanoscience - MC2, Chalmers University of Technology, 412 96 G\"oteborg, Sweden \\
$^{2}$ Laboratoire de Physique de l'\'Ecole Normale Sup\'erieure, ENS, Universit\'e PSL, CNRS, Sorbonne Universit\'e, Universit\'e de Paris, 75005 Paris, France}

%\author{ B. Rousseaux    }
%\affiliation{Laboratoire de Physique de l'\'Ecole Normale Sup\'erieure, ENS, Universit\'e PSL, CNRS, Sorbonne Universit\'e, Universit\'e de Paris, F-75005 Paris, France}

%\author{ G. Johansson    }
%\affiliation{Applied Quantum Physics Laboratory, Department of Microtechnology and Nanoscience - MC2, Chalmers University of Technology, 412 96 G\"oteborg, Sweden}
%\author{}
%\thanks{}
%\affiliation{}


\date{\today}

\begin{abstract} 
	In this letter, we investigate the dynamics of a single superconducting artificial atom capacitively coupled to a transmission line with a characteristic impedance comparable or larger than the quantum resistance. In this regime, microwaves are reflected from the atom also at frequencies far from the atom's transition frequency.  Adding a single mirror in the transmission line then creates cavity modes between the atom and the mirror. Investigating the spontaneous emission from the atom, we then find Rabi oscillations, where the energy oscillates between the atom and one of the cavity modes. 
\end{abstract}

%\keywords{}

\maketitle

%\section{}
%\subsection{}
%\subsubsection{}
\section{Introduction}
In the past two decades, circuit quantum electrodynamics (circuit QED) has become a field of growing interest for quantum information processing and also to realize new regimes in quantum optics \cite{Roy2017, Gu2017,GonzCira17_2,PaulGonz18,QuijJoha18,Wallraff2004,Blais2004,Wendin2017,Gu2017,Kockum2019}. The restriction to one dimensional (1D) waveguides  in circuit and waveguide QED enhances directionality and reduces losses and therefore has a great advantage over higher dimensional systems to reach strong- and ultrastrong-coupling regimes \cite{ZhenBara10,Tufarelli2013,Tufarelli2014,SancMart14,SancZuec16,SancGarc17,DevoScho07,BambOgaw14,USC_Anton,ZuecGarc19,MlynWall14}. 
A typical circuit QED set-up consists of a superconducting qubit coupled to a 1D transmission line (TL) \cite{Wallraff2004,Blais2004,Wendin2017,Gu2017,Pozar}. Superconducting qubits are artificial atoms built with a non-linear Josephson Junction (JJ), that creates an anharmonic energy spectrum \cite{Kockum2019}. There are different kinds of superconducting qubits like flux qubits, phase qubits, and charge qubits \cite{Wendin2017,Clarke2008}.
A 1D transmission line can be modelled by coupled LC oscillators, each having a characteristic impedance of $Z_0 = \sqrt{L_0/C_0} \approx 50\,\si{\Omega}$, smaller than the quantum resistance $R_Q = \hbar/(2e)^2 \approx 1.0 \, \si{k \Omega}$. But recent studies show that it is possible to reach impedances comparable to the quantum resistance or higher \cite{Pechenezhskiy2020,Niepce2019,Masluk2012,Weissl2015,Samkharadze2016,Kuzmin2019,PuertasMartinez2019}. This can be realized by building circuits made of arrays of JJs \cite{Bell2012,Winkel2020,Krupko2018,Masluk2012,Weissl2015,Kuzmin2019,PuertasMartinez2019} or high-kinetic-inductances materials, called superinductors \cite{Peltonen2018,Grunhaupt2018,Samkharadze2016,Niepce2019,Niepce2020,Pechenezhskiy2019}. High impedance JJ arrays and superinductors are for example used in the Fluxonium qubit \cite{Manucharyan2009,Gruenhaupt2019,Nguyen2019,Koch2009}, which has reduced charge noise sensitivity and can have relaxation times up to milliseconds \cite{Pop2014,Nguyen2019}.
This also has an advantage for metrology, since the charge noise insensitivity makes it possible to measure the current very accurately \cite{Guichard2010}.
Furthermore, high-impedance resonators make it possible for light-matter interaction to reach strong coupling regimes, due to strong coupling to vacuum fluctuations \cite{Stockklauser2017}.\\
In this article, we investigate the spontaneous emission of a transmon \cite{Koch2007} capacitively coupled to a 1D TL that is shorted at one end. This system is known as an "atom in front of a mirror" \cite{Hoi2015,Wen2018,Wen2019,Peng2016,FornDiaz2017}. Instead of using a Markovian master equation approach, we are taking the photon traveling time fully into account, making the dynamics non-trivial 
%\brcolour{(The dynamics of this system is usually understood in terms of a Markovian master equation approach, with the cost of neglecting the time delay of a single photon travelling between the atom and the mirror. Here, we exploit a semi-classical set of linearized differential equations with a delay, revealing non-trivial dynamical effects.)}
\cite{Dorner2002,Guo2017,Tufarelli2014,Bradford2013,Grimsmo2015,Pichler2016,Pichler2017,Guimond_2017,SchnBusc16,GonzCira17,Wiegand2020}. Furthermore, we explore the above-mentioned regime of a TL impedance exceeding the quantum resistance $Z_0 \gg R_Q$. We find that the system behaves qualitatively different, compared to the well studied low impedance regime $Z_0 \ll R_Q$. The atom reflects strongly at all frequencies, except its transition frequency.  Together with the mirror, it thus forms a cavity and when the transition frequency is close to a cavity mode, we find a vacuum Rabi-splitting, resulting in Rabi oscillations in the spontaneous emission. In this regime, all dynamical timescales are independent of the coupling capacitance and instead depend on the intrinsic transmon capacitance and the TL impedance.
Another cavity-free system that shows Rabi splitting is an artificial atom coupled surface acoustic waves \cite{Ask2019}.
\section{Circuit-QED model}
\label{circuitQED}

\begin{figure}[t]
	\centering
	\begin{minipage}{1\linewidth}
		\begin{overpic}[width=1\linewidth]{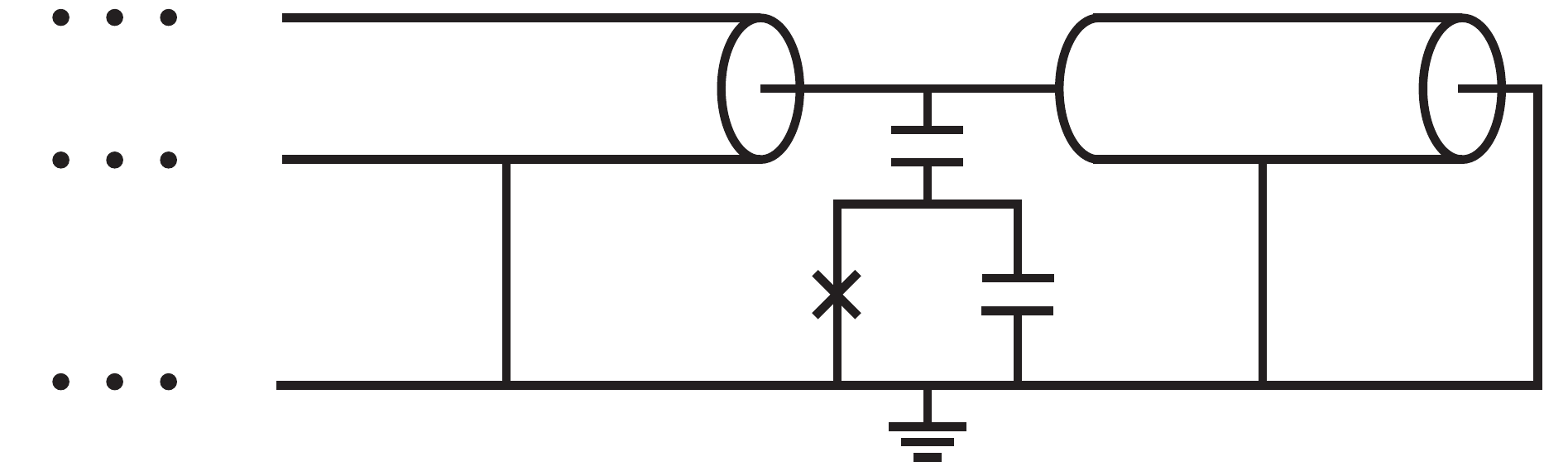}
			\put(51,20){ $C_c$}
			\put(46,17){ $\phi_J$}
			\put(56,27){ $\phi_0$}
			\put(45,10){ $E_J$}
			\put(69,10){ $C_J$}
			\put(30,23){ $Z_0$}
			\put(78,23){ $Z_0$}
			\put(-3,23){ $a)$}
			\put(-3,-6){ $b)$}
			%	\begin{overpic}[width=1\linewidth]{Sketches/Circuit.png}
			%		\put(18,21){$\Phi_{-1}$}
			%		\put(40,21){$\Phi_{0}$}
			%		\put(60,21){$\Phi_{1}$}
			%		\put(33,15){ $C_C$}
			%		\put(42,12.5){ $\Phi_j$}
			%		\put(31,8){ $E_J$}
			%		\put(45,8){ $C_J$}
			%		\put(5,22){ $\Delta x L_0$}
			%		\put(24,22){ $\Delta x L_0$}
			%		\put(46.5,22){ $\Delta x L_0$}
			%		\put(65,22){ $\Delta x L_0$}
			%		\put(87,22){ $\Delta x L_0$}
			%		\put(7,10.5){ $\Delta x C_0$}
			%		\put(63,10.5){ $\Delta x C_0$}
			%		\put(45,8){ $C_J$}
			%		\put(-1,20){ $a)$}
			%		\put(-1,-3){ $b)$}
			
			%\visible<2->{\put(41,30){Connection?}}
		\end{overpic}
	\end{minipage}
	\begin{minipage}{1\linewidth}
		\vspace{0.2cm}
		\begin{overpic}[width=1\linewidth]{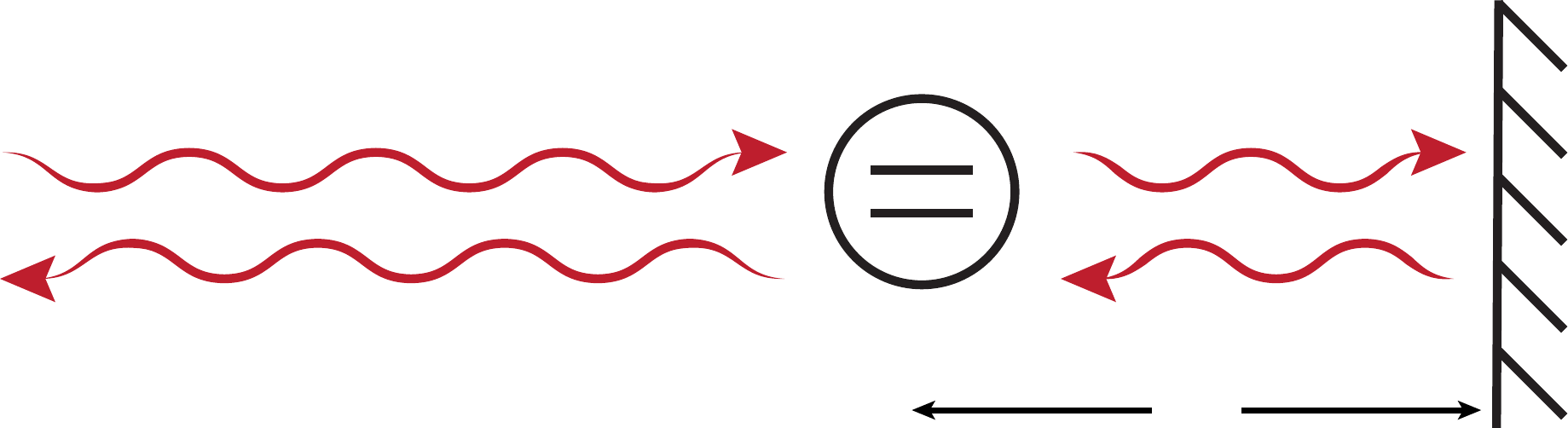}
			\put(75,21){$\VRout$}
			\put(30,21){$\VLin$}
			\put(75,5){$\VRin$}
			\put(30,5){$\VLout$}
			\put(75,0){$L$}
		\end{overpic}
	\end{minipage}
	\caption{a) The circuit model of a transmon coupled through the coupling capacitance $C_c$ to a semi-infinite 1D TL with impedance $Z_0$.  The  Josephson energy, flux, and capacitance of the transmon are denoted by $E_J$, $\phi_J$, and $C_J$. The flux on the coupling capacitance $C_c$ is denoted by $\phi_0$ with the corresponding voltage $V_0 = \dot \phi_0$.\\
		b) A sketch of the system depicting an atom in front of a mirror coupled to incoming/outgoing microwave fields to the left/right characterised by their respective voltages $V_{L/R}^{\text{in/out}}$ at the transmon. The mirror couples the fields to the right $\VRin(t)=-\VRout(t-T)$, introducing the time of propagation to the mirror and back $T=2L/v$.
	}
	\label{fig:System}
\end{figure}
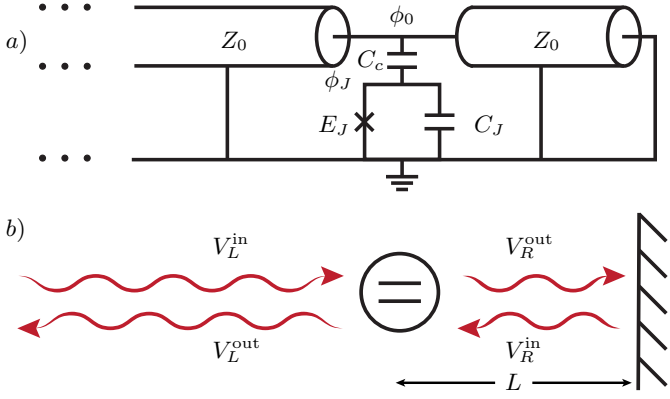
Our system consists of a transmon qubit capacitively coupled to a semi-infinite 1D TL at a distance $L$ from its grounded end (see Fig.~\ref{fig:System}). The transmon qubit consists of a superconducting anharmonic $LC$-oscillator, where the inductive ($L$) element is formed by a Josephson junction (JJ) with characteristic energy $E_J$ in parallel with a capacitor ($C$) with capacitance $C_J$. The sinusoidal current-phase relation of the JJ makes the energy spectrum of the transmon qubit anharmonic, allowing for excitation with a single microwave photon using standard harmonic microwave sources. The transmon is capacitively coupled to a microwave TL, characterised by its inductance per unit length $L_0$ and capacitance per unit length $C_0$. The velocity and impedance of the electromagnetic field inside the TL is given by $v= 1/\sqrt{L_0 C_0}$ and $Z_0=\sqrt{L_0/C_0}$, respectively. 
%The grounded node at $N=L/\Delta x$ reflects the electromagnetic field with a round-trip time of $T=2v/L$. A sketch of the system is depicted in Fig.~\ref{fig:System}.
% The electromagnetic field in the TL is reflected at the grounded end at the node $N=L/\Delta x$, with the distance $L$ to the transmon which is connected to the node $N = 0$. The system can be described by an atom in front of the mirror. A sketch of the lumped element circuit model and the system as an atom in front of a mirror with an electromagnetic field is depicted in Fig.~\ref{fig:System}. 
%We derive the Hamiltonian of the lumped element model 
Using the standard circuit quantization procedure \cite{Devoret1995,Wiegand2020}, we can derive
%with our system coordinates given by the node fluxes $\phi_i (t) = \int_{0}^{t} V_i (t') dt'$ with the voltage $V_i$ over node $i$, and the conjugate variable $p(t)$, which fulfil the canonical commutation relations $\left[ \phi_i, p_j\right] = i \hbar \delta_{ij}, \left[\phi_i, \phi_j\right] = \left[p_i, p_j\right] = 0$. The 
the Heisenberg equations of motion for the charge $p_0(t)$ on the coupling capacitor $C_c$, the charge $p_J(t)$ on $C_J$ and its conjugate flux $\phi_J (t)$, giving the phase difference $2e \phi_J (t)/\hbar$ over the JJ. Denoting the operators for the voltages of the incoming/outgoing microwave fields to the left/right of the transmon as $V_{L/R}^{\text{in/out}}(t)$ (see Fig.~\ref{fig:System}b) these equations are   
%at the coupling point $x=0$ 
%in the continuum limit $\Delta x \rightarrow 0$ are given by
%
%
%

\begin{align}
\partial_t \phi_J (t) &= \frac{1}{C_J} (p_J(t)  + p_0(t) ), \label{eq:EoMphiJa}\\
	\partial_t p_J (t) &= - E_J \frac{2 e}{\hbar} \sin ( \frac{2 e}{\hbar}\phi_J(t) ) \approx -\frac{\phi_J(t)}{L_J}, \label{eq:EoMpJa} \\
	\partial_t p_0(t)  &= \frac{2 p_0(t)}{Z_0 C_\Sigma}  + \frac{2 p_J(t)}{Z_0 C_J}  - \frac{2}{Z_0} \left[ V_L^{\text{in}}(t)+V_R^{\text{in}}(t)\right]   , \label{eq:EoMp0OpenTL}\\
	V_{L/R}^{\text{out}}(t) &= V_{R/L}^{\text{in}}(t) - \frac{Z_0}{2} \partial_t p_0(t), \label{eq:V}
\end{align} 
where we denoted the capacitance to ground seen by the JJ as $C_\Sigma = C_c C_J / (C_c + C_J)$ and in the second equation, we introduced the Josephson inductance $L_J =\hbar^2/4e^2E_J$, which describes the linearized dynamics of the Josephson junction. This approximation is obviously good in the weak excitation regime $|\phi_J(t)| < \hbar/2e$ and will also be sufficient to describe the spontaneous emission, where the transmon is initially excited by a single microwave photon.% \textcolor{red}{What about reflection?}
%Here, we replaced the discrete coordinates $\phi_i (t)$ by a continuous flux field $\phi (x, t)$ with $x_i = i \Delta x$, with the Transmon located at $x=0$. Similarly, we replaced the TL charges $p_i (t)$ by a charge density field  $p\left(x_{i}, t\right)=p_{i}(t) / \Delta x$. 

%Here, the variables are the semi-classical averages of the coordinates and the qubit was linearised by extending the cosine potential to second order \cite{Wiegand2020}. 

%\textcolor{red}{Maybe write something about the coupling and summarize the old paper, so that all the important things are already introduced.}
%
%
%
%
%To be able to analyse the dynamics of the transmon regarding the TL impedance $Z_0$, we define the characteristic impedance of the transmon
%
%
%\begin{align}
%Z_J = \sqrt{\frac{L_J}{C_J}} = R_Q \sqrt{\frac{2 E_C}{E_J}},
%\end{align}
%
%
%where $E_C = e^2/(2C_J)$ is the charging energy of the transmon. In the following, we analyse different energies of the system for different ratios of the TL and Transmon impedances $Z_0 / Z_J$. \textcolor{red}{Maybe say something more here about the coupling?}
%In the following, we will investigate the reflection of the transmon if a probe field is applied from one side.
%
%
%
\section{Reflection}

\subsection{Open TL}
To characterise how the behaviour of the system changes when we increase the TL impedance $Z_0$, we first investigate the reflection of microwaves from the transmon coupled to an open TL, i.e. without a mirror.
Since the equations of motion are linear, we can express the reflected field operator $\VLout$ in terms of the incoming probe field operator $\VLin$ by Fourier transforming the equations of motion \eqref{eq:EoMphiJa}-\eqref{eq:V}, assuming no incoming field from the right ($\VRin=0$). The expression for the frequency dependent reflection coefficient is given by

%The reflection coefficient of a transmon coupled to an open transmission line that is probed from the left is given by $r = \VLout / \VLin$, where $\VLin$ is the amplitude of the incomin field from the left and $\VLout$ is the amplitude of the reflected field. A sketch of the electromagnetic field in the TL is depicted in Fig.~\ref{fig:System} b). By using the wave representation of the variables \cite{Wiegand2020}, we calculate the reflection coefficient by Fourier transforming and solving the equations of motion \eqref{eq:EoM1}-\eqref{eq:EoM3}, where we neglect the time-delay term.  The reflection is given by
%
%
\begin{figure}[t]
	\centering
	\includegraphics[width=1\linewidth]{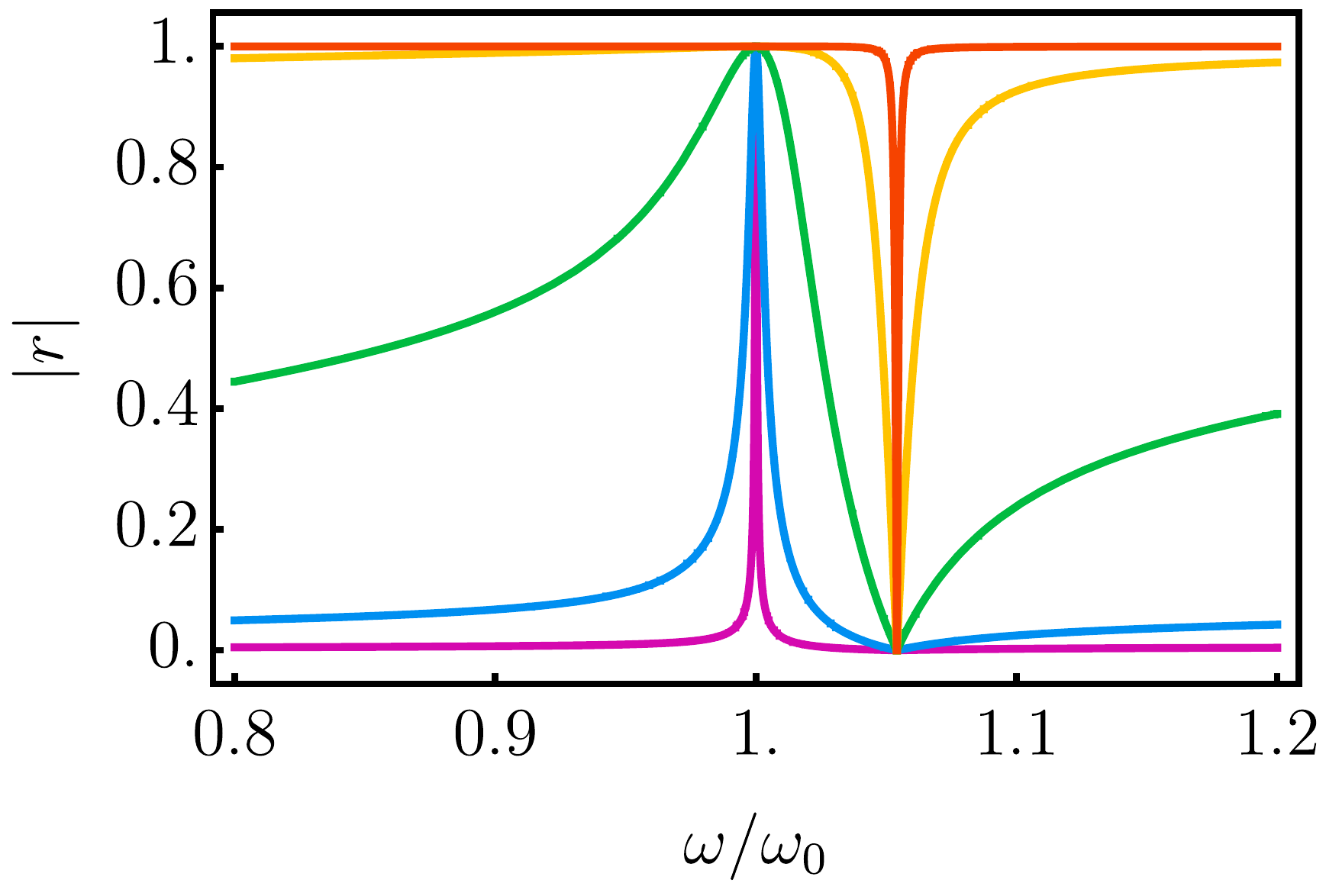} %HighImpedance.nb
	\caption{Reflection of a transmon in an open TL for different ratios of the TL and qubit impedance $Z_0/Z_J$. The curves show the reflection for $\frac{C_c}{C_c+C_J} = 0.1$ and  $Z_0/Z_J = 0.1$ (purple), $Z_0/Z_J = 1$ (blue), $Z_0/Z_J = 10$ (green), $Z_0/Z_J = 100$ (yellow), and $Z_0/Z_J = 1000$ (red) from low to high  TL impedance. For low impedance, the qubit reflects only at the coupled qubit frequency $\omega_0$, but for high impedance, it reflects everywhere but the uncoupled qubit frequency $\omega_J$. 
	}
	\label{fig:Reflection}
\end{figure}
\begin{align}
r(\omega) \equiv \frac{\VLout(\omega)}{\VLin(\omega)} = \frac{C_c Z_0 \omega \left( \frac{\omega^2}{\omega_J^2} - 1 \right) }{  2 i \left( 1 - \frac{\omega^2}{\omega_0^2}\right) + C_c Z_0 \omega \left( \frac{\omega^2}{\omega_J^2} - 1 \right)}
\label{eq:r_open_line}
\end{align}
where $\omega_0 = 1/\sqrt{L_J (C_c + C_J)}$ is the resonance frequency of the coupled transmon and $\omega_J = 1/\sqrt{L_J C_J}$ is the resonance frequency of the bare (uncoupled) transmon. In Fig.~\ref{fig:Reflection} the reflection around the transmon resonance frequencies is shown for different values of $Z_0$. We see that for low impedance $Z_0 C_c \omega < 1$, the reflection is weak except at $\omega_0$ where it is unity, due to resonant reflection from the transmon \cite{Lindkvist2013}. For high impedance $Z_0 C_c \omega > 1$ we instead see strong reflection at all frequencies, except around the "new" resonance frequency $\omega_J$,  where we find zero reflection independent of $Z_0$. The crossover occurs at $Z_0 \sim Z_J C_J/C_c$, introducing the transmon impedance  
\begin{align}
Z_J = \sqrt{\frac{L_J}{C_J}} = R_Q \sqrt{\frac{2 E_C}{E_J}}, 
\end{align}
where $E_C = e^2/(2C_J)$ is the charging energy of the transmon.

In the high impedance regime, the strong scattering away from $\omega_J$ occurs due to the comparably strong capacitive coupling to ground at the transmon {\em without exciting the transmon}. Close to $\omega_J$, the resonantly excited transmon counteracts this capacitive coupling and effectively acts like an open circuit. This is the opposite behaviour compared to the low impedance regime, where the transmon is effectively an open circuit at all frequencies, except at its resonance frequency $\omega_0$, where it acts like a shorted circuit, giving full reflection. By fitting Lorentzians to Eq.~\eqref{eq:r_open_line}, we can extract the high impedance coupling strength $\gamma_J=2/Z_0 C_J$, which in contrast to the low-impedance expression $\gamma_0 = Z_0 C_c^2 /2 L_J (C_J+C_c)^2$ does not depend on $C_c$ and decreases with increasing $Z_0$. In this regime, the voltage $V_0=\VLin+\VLout=\VRin+\VRout$ at the node coupling to the TL oscillates with the full voltage across the JJ, $V_J=\dot{\phi_J}$. Due to the large $Z_0$, the amount of current passing through the TL is small ($\propto V_J/Z_0$), thus not being able to change the voltage $V_0-V_J$ across $C_c$ significantly, i.e. $|V_0-V_J| \ll V_J$. In the low impedance regime, we instead have large currents flowing through the TL, keeping the voltage at the coupling node close to zero, i.e. $|V_0| \ll |V_J|$. These currents obviously scale with $C_c$ and the energy dissipation scales with $Z_0$. 
%
%
%%
%\begin{align}
%r = \left\{\begin{array}{llll}
%1, &  \omega = \omega_0  &\forall &Z_0\\
%0, &   \omega = \omega_J &\forall &Z_0
%\end{array}\right. ,
%\end{align}
%%
%and the limits of this function are given by
%%
%%
%%
%\begin{align}
%\lim_{Z_0 \to \infty} r =\left\{\begin{array}{ll}
%1, &  \omega \neq \omega_J \\
%0, & \omega = \omega_J
%\end{array}\right. , \hspace{0.2cm}
%\lim_{Z_0 \to 0} r =\left\{\begin{array}{ll}
%1, &  \omega = \omega_0 \\
%0, &   \omega \neq \omega_0
%\end{array}\right.,
%\end{align}
%%
%%
%where the cross-over occurs at $Z_J/Z_0 \sim %C_c/C_J$ which explains the behaviour seen in %Fig.~\ref{fig:Reflection}. 
%%
%%
%This shows that the system drastically changes by increasing the ratio of the TL impedance to the qubit impedance $Z_J/Z_0$. For low impedance, the reflection shows a Lorentzian curve around the coupled qubit frequency $\omega_0$, which is the behaviour we expect \cite{Lindkvist2013}. For very high impedance, the qubit reflects for all frequencies except the uncoupled qubit frequency $\omega_J$. This means the resonance frequency changes, depending on the ratio $Z_0/Z_J$. 
In the following, we investigate how the mirror affects the scattering.
\subsection{Mirror}
\begin{figure}[t]
	\centering
		\begin{overpic}[width=1\linewidth]{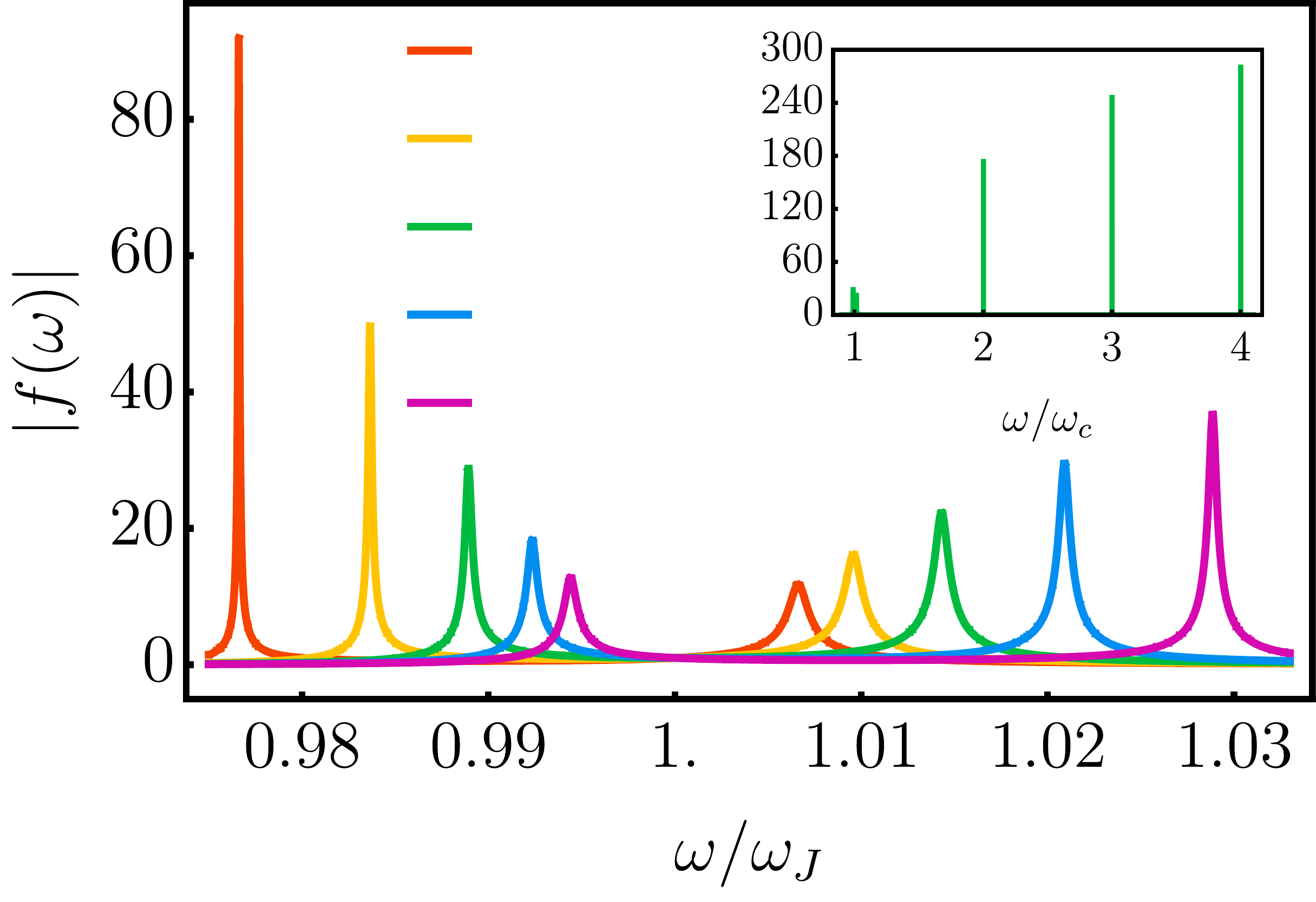} %ReflectionMirror.nb
			\put(37,63.7){$\frac{\omega_c}{\omega_J} = 0.98$}
			\put(37,57){$\frac{\omega_c}{\omega_J} = 0.99$}
			\put(37,50.3){$\frac{\omega_c}{\omega_J} = 1$}
			\put(37,43.7){$\frac{\omega_c}{\omega_J} = 1.01$}
			\put(37,36.9){$\frac{\omega_c}{\omega_J} = 1.02$}
			\put(36,32){$\xleftarrow{\hspace*{1.1cm}}$}
	        \put(56.5,32){$\xrightarrow{\hspace*{1.1cm}}$}
	        \put(52.5,32){$\Omega$}
		\end{overpic}
	\caption{  Amplitude of the electromagnetic field between the qubit and the mirror $|f(\omega)| =  \left| \VRout(\omega)/\VLin(\omega) \right|$ for $\frac{C_c}{C_J} = 0.1$ and  $Z_0/Z_J = 1000$. The delay time is given by $T = 2 \pi n /\omega_c, n=1$ and $\omega_c$ varies with respect to $\omega_J$. The different colors show $\omega_c/\omega_J = 0.98$ (red), $\omega_c/\omega_J = 0.99$ (yellow), $\omega_c/\omega_J = 1$ (green), $\omega_c/\omega_J = 1.01$ (blue), $\omega_c/\omega_J = 1.02$ (purple). All curves, show a splitting which can be understood as a Rabi splitting. Without detuning between $\omega_J$ and $\omega_c$, the splitting is centered exactly around $\omega/\omega_J = 1$. The inset shows the appearance of higher cavity modes appearing at $\omega \approx n \omega_c$, where  $\omega_c T= 2 \pi$ for $\omega_c/\omega_J = 1$ corresponding to the green curve in the main figure.
	}
	\label{fig:ReflectionMirror}
\end{figure}
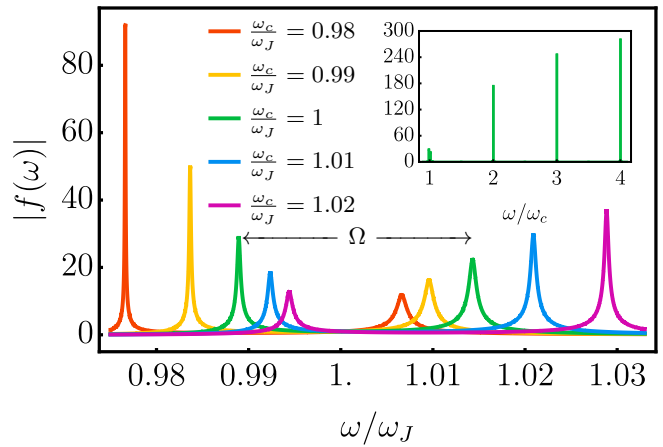
%\begin{figure}[t]
%	\centering
%	\includegraphics[width=1\linewidth]{ReflMirrorNode097Inset.pdf} %ReflectionMirror.nb
%	\caption{ Amplitude of the electromagnetic field between the qubit and the mirror for $\frac{C_c}{C_c+C_J} = 0.1$ and  $Z_0/Z_J = 10$ (purple), $Z_0/Z_J = 50$ (blue), $Z_0/Z_J = 100$ (green), $Z_0/Z_J = 300$ (yellow), and $Z_0/Z_J = 1000$ (red). The delay time here is given by $T = 2 \pi n /\omega_0, n=0.97$, implying that the qubit is located slightly closer to the mirror than the first node of the electromagnetic field. The inset shows the appearance of cavity modes appearing at $\omega \approx n \omega_c$, where  $\omega_c T= 2 \pi$ for $Z_0/Z_J = 50$.
%	}
%	\label{fig:ReflectionMirror}
%\end{figure}
%
The mirror couples the fields to the right of the transmon $\VRin(t)=-\VRout(t-T)$, introducing the time of propagation to the mirror and back $T=2L/v$, see Fig~\ref{fig:System} b). Similarly as before, we can find the response to a harmonic field incoming from the left by Fourier transformation of the equations of motion.
%but this time with the time-delay included. This means that since field is reflected at the mirror, the ingoing field from the right side $\VRin$ is the phase shifted outgoing field $\VRout$, $\VRin = - \VRout \e^{i\omega T}$. 
Since the absolute value of the reflection for the transmon in front of a mirror is always unity, we are now interested in the frequency dependence of the ratio between the trapped field (between the qubit and the mirror) and the incoming field, which is given by $f(\omega) \equiv \VRout(\omega)/\VLin(\omega)$,
\begin{align}
\label{f_omega}
	f(\omega)= \frac{ \left( \frac{\omega^2}{\omega_0^2} - 1 \right) }{    \left( 1 - \frac{\omega^2}{\omega_0^2}\right) - i \frac{C_c Z_0 \omega }{2} \left( \frac{\omega^2}{\omega_J^2} - 1 \right) \left( e^{i \omega T} - 1  \right) },
\end{align}
which is shown in Fig.~\ref{fig:ReflectionMirror}.
In the high impedance regime, we now find cavity resonances between the highly reflective atom and the mirror when the frequency is close to $n\ \omega_c$ for $n=1,2,\dots$ and $\omega_c = 2 \pi/T$, as shown by the peaks in the inset of Fig.~\ref{fig:ReflectionMirror}. These are broadened by the coupling to the TL by $\gamma_c^n=|t(n\ \omega_c)|^2/T$, where $t(\omega)$ is the transmission across the transmon and $|t(\omega)|^2=1-|r(\omega)|^2$. 
%$\gamma_c^n=\left(\pi n^2 C_c^2 Z_0^2 \omega_c \right)^{-1}$. 
We find that the effect on the transmon resonance close to $\omega_J$ is simply to reduce its broadening with a factor of two to $\gamma_J^m = 1/Z_0 C_J$, away from any qubit-cavity resonance $\omega_J \approx n\ \omega_c$. 
%$|f(\omega)|^2$ also gives the ratio between spectral density of vacuum voltage fluctuations in the region to the right of the transmon compared to the flat density in the open transmission line. 
As shown in the main panel of Fig.~\ref{fig:ReflectionMirror}, on resonance $\omega_J \approx n\ \omega_c$ we find an avoided crossing with the coupling strength 
\begin{align}
    \Omega &= \frac{2}{ \sqrt{T C_J Z_0 } } 
 =\frac{2\omega_J}{ \sqrt{2 \pi n C_J Z_0 \omega_J} } 
    = \frac{ 2 \omega_J}{ \sqrt{2 \pi n  \frac{Z_0}{Z_J} } }.
\label{eq:omegaB}
\end{align}
As we will see in the next section, where we investigate the spontaneous emission dynamics of the transmon, we find that this coupling indeed gives rise to vacuum Rabi-oscillations between the transmon and the cavity mode.

In the low impedance regime $Z_0 C_c \omega < 1$, $f(\omega)$ is instead close to unity, indicating only little scattering from the transmon for all frequencies far from the transmon resonance $\omega = \omega_0$. Here, $f(\omega_0)=0$, since the field is reflected by the transmon and does not reach the mirror. When the transmon is located at a distance corresponding to a node of the electromagnetic field at its resonance frequency, i.e. $\omega_0 T = 2 n \pi$, it is in a dark state and is thus completely invisible to the incoming field at frequency $\omega_0$, giving instead $f(\omega_0)=1$. In the dark state, both the transmon and the field between the transmon and the mirror are excited. Thus, if the distance is slightly longer/shorter than the node, the state is no longer completely dark, and we instead get a pronounced scattering resonance ($|f(\omega)|\gg 1$) at frequencies slightly lower/higher than $\omega_0$, see e.g. the purple line in Fig.~\ref{fig:ReflectionMirror}. For higher $Z_0$, we see that this dark state resonance moves in frequency towards the cavity frequency $\omega_c = 2 \pi/T$. As shown in the supplemental material, for small $C_c/C_J \ll 1$, and $\omega_0=n\ \omega_c$ we can find vacuum Rabi-oscillations damped towards a finite dark state population.  
%, which we can straightforwardly understand as a cavity resonance, since the transmon now forms a highly reflective mirror at all frequencies except $\omega_J$. In the inset of this figure, we indeed see that cavity resonances appear at $n\ \omega_c$ for $n=1,2,\dots$ as expected.

%\textcolor{blue}{In addition, we see less pronounced scattering resonance appear close to $\omega_J$. For high $Z_0$, the scattering phase of the transmon changes rapidly around this frequency and the resonance condition for a pronounced standing wave between the transmon and the mirror can be met for almost any positive distance $T>0$.}
%
 %In Fig. \ref{fig:ReflectionMirror}, we plot this normalized amplitude of the electromagnetic field between the qubit and the mirror for different values of $Z_0/Z_J$. We see two kinds of resonances: a resonance close to $\omega_0$ and a resonance at $\omega_J$ which is sharp for high impedance (red curve) and becomes flatter as the TL impedance decreases (purple curve). The inset shows the field amplitude for higher modes for $Z_0/Z_J=50$. This indicates that especially for high impedance, the qubit behaves like a mirror and creates a cavity together with the shorted end of the TL with cavity modes $\omega_c$. 
 In the following, we investigate how the high-impedance TL influences spontaneous emission of the transmon.
\section{Spontaneous emission and Rabi oscillations}
\begin{figure}[t]
	\centering
	\includegraphics[width=1\linewidth]{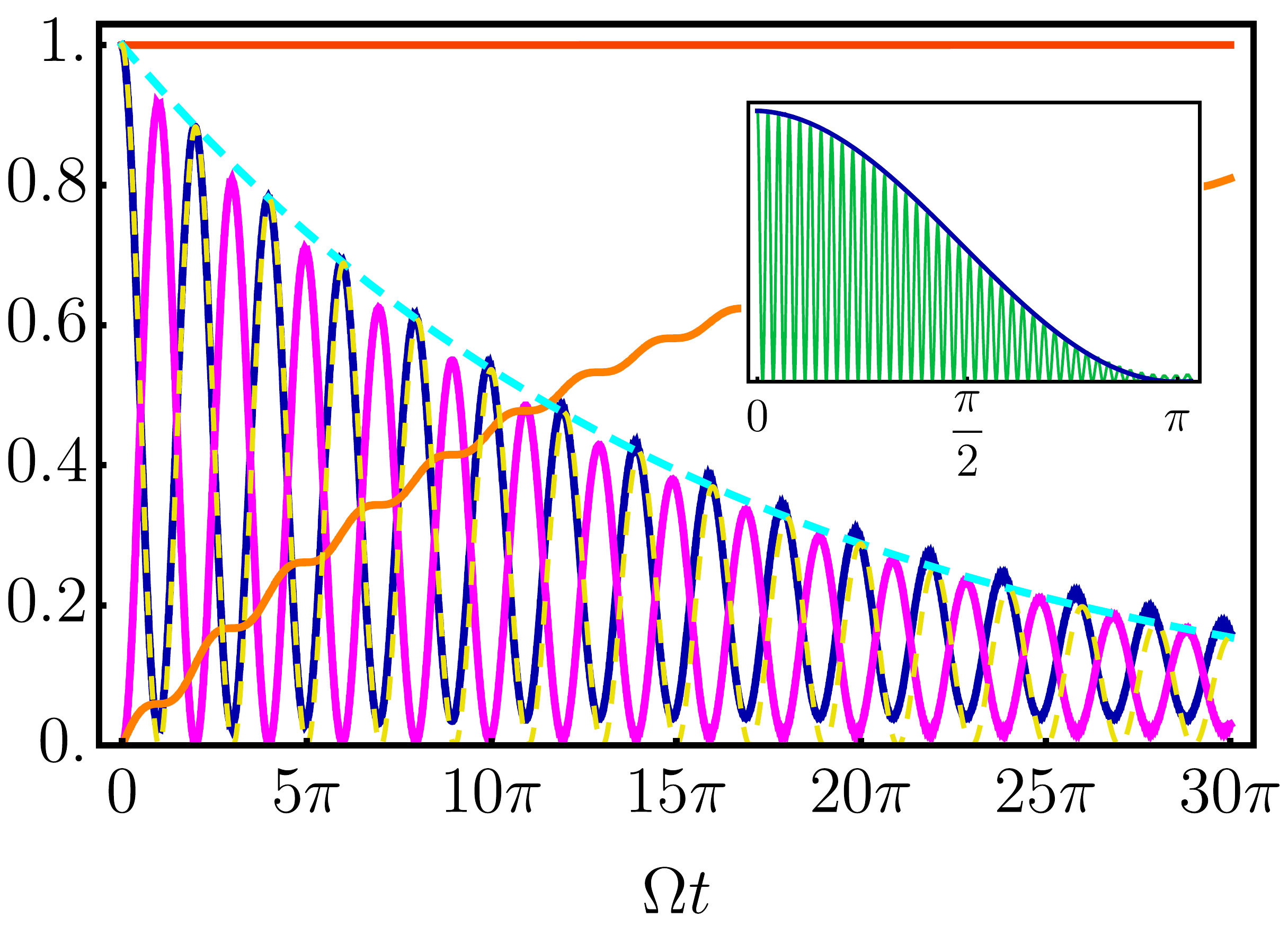} %HighImpedance.nb
	\caption{Energy of the transmon (blue), Field between the mirror and transmon (pink), outgoing field to the left (orange), sum of all the energies (red), approximated decay $\gamma_J^m$ (cyan, dashed) and approximated Rabi frequency $\Omega$ and decay $e^{-\gamma^m_J t/2} \cos^2 (\Omega t/2)$ (yellow, dashed) for $C_c/C_J = 0.1, Z_0/Z_J = 1000, T = 2 \pi/\omega_C, \omega_C = \omega_J$ as a function of periods of the Rabi oscillations. The Inset shows the energy of the transmon (blue) and the flux $\phi_J^2$ (green) for the first oscillation of the frequency $\Omega$. This clearly shows the different timescales of the energy compared to the variables of the transmon.
	}
\label{fig:Energies}
\end{figure}
We consider the case of a transmon initially excited at time $t=0$ with a finite flux $\phi_J(0)>0$, while the other qubit variables are zero $p_J(0)=p_0(0)=0$ and the TL is in the vacuum state. The qubit energy
\begin{equation}
E_q(t)=\frac{\left(p_J(t)+p_0(t)\right)^2}{2C_J}+\frac{p_0(t)^2}{2C_c} + \frac{\phi_J(t)^2}{2L_J},
\end{equation} 
is the sum of the capacitive energy on the two capacitances and the inductive energy in the JJ. The current amplitude emitted from the transmon into the TL is $\partial_t p_0(t)$ and from this we can write the change of the energy $E_R(t)$ of the field between the transmon and the mirror as
\begin{equation}
\partial_t E_R(t) = \frac{Z_0}{4} \left[ (\partial_t p_0(t))^2 - (\partial_t p_0(t-T))^2 \right],
\end{equation}
where the first term corresponds to the instantaneous power emitted into the TL and the second term is the instantaneous power coming back from the mirror. The change of the energy of the field to the left of the transmon $E_L(t)$ is given by the instantaneous left-moving power leaving the system,
\begin{equation}
\partial_t E_L(t) = \frac{Z_0}{4} \left[ \partial_t p_0(t) - \partial_t p_0(t-T) \right]^2,
\end{equation}
where the left-moving current amplitude is a sum of the current emitted by the transmon and the delayed current arriving from the mirror.

%The energy on the right side of the qubit $E_R$, is the energy that is trapped between the qubit and the mirror. The energy on the left side of the qubit $E_L$, is the energy that goes into the open end and leaves the system. 
In Fig.~\ref{fig:Energies}, we plot these energies for $Z_0/Z_J=1000$ for the case of resonance between transmon and the first cavity resonance $\omega_J=\omega_c$. The system energies indeed perform damped Rabi-oscillations with the frequency $\Omega=2/\sqrt{T C_J Z_0}$ and half the off-resonance damping rate $\gamma_J^m/2=1/2C_J Z0$, as indicated by the yellow line given by the expression $e^{-\gamma^m_J t/2} \cos^2 (\Omega t/2)$, approximating the full numerical solution of the differential equations very well. We note that Laplace transforming the equations of motion \eqref{eq:EoMphiJa}-\eqref{eq:V} and calculating the residues of the system variables, gives similar expressions for the Rabi frequency and damping rate as the analysis of the resonances in the scattering amplitudes. More details and a comparison of the approximation to the numerical results can be seen in the supplemental materials.

%we are able to find the frequency of these oscillations, which we find to be very close to the system frequencies which results in a beating effect, see Fig. \ref{fig:Energies}. For high TL impedance, we are able to find an approximation for the Rabi oscillation by analyzing the poles of the Laplace transformed system variables. The approximated Rabi frequency is given by
%\brcolour{\{Could we compare this with $\omega_c - \omega_J$? If this is say, Rabi oscillations between the cavity and the JJ, then I expect that this beating would be like $\omega_B = \omega_c-\omega_J$. Looking at Fig. 4, I also notice that the beating strikingly looks like \emph{resonant} Rabi oscillations, suggesting that you actually have some kind of strong coupling between two resonant (unidentified?) modes, yielding the double peak spectrum seen in Fig. 3 for high $Z_0$.\}}
%
%
%%
%
%For this approximation we assumed that $\frac{C_c}{C_J} \frac{Z_0}{Z_J}> 1$. In the supplemental material, we demonstrate how the analytically and numerically calculated beating frequencies deviate from each other. We find that they become closer the higher $\frac{C_c}{C_J} \frac{Z_0}{Z_J}$ becomes. Similarly, we can find an approximation for the decay, $\gamma_J = 1/(2 C_J Z_0)$, which is depicted as the green curve in Fig~\ref{fig:Energies}. The yellow curve shows both approximations, $e^{\gamma_J t} \cos (\Omega_n t)$ and we see that they match the numerical data well.
%
%
%
%
%
\section{Effective quantum model: atom in a multimode cavity Hamiltonian}
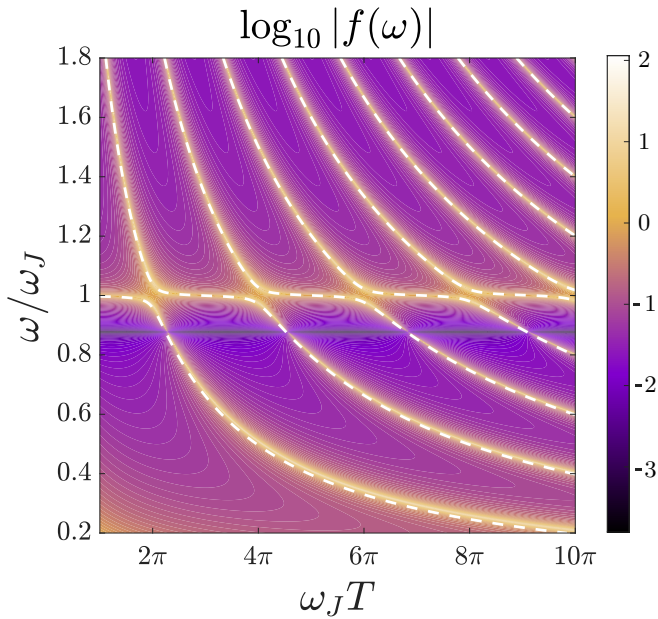
\begin{figure}
	\centering
	\begin{overpic}[width=1\linewidth]{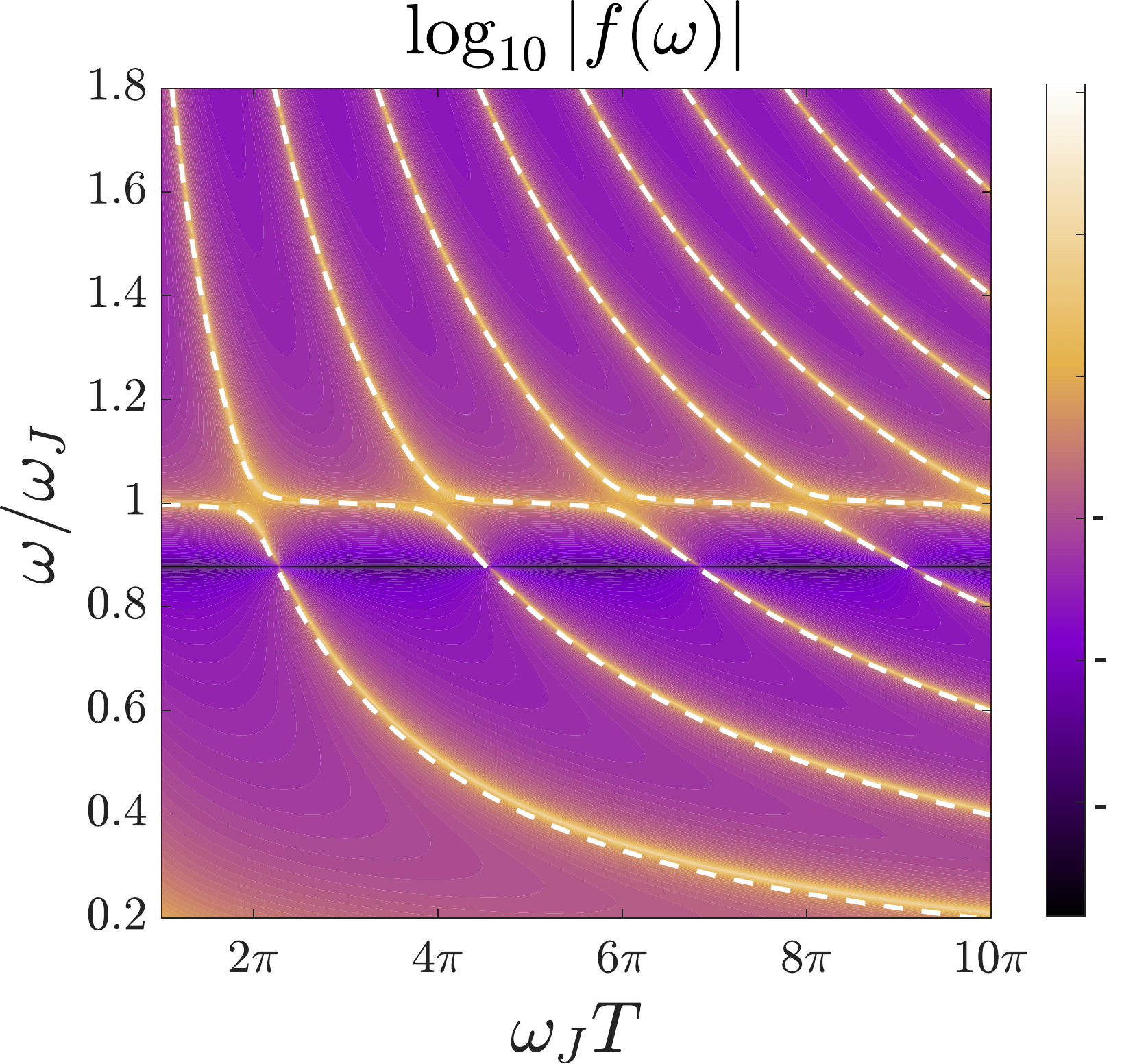}
		\put(98,21.53){3}
		\put(98,34.5){2}
		\put(98,47){1}
		\put(97,59.4){0}
		\put(97,72){1}
		\put(97,84.5){2}
	\end{overpic}
	\caption{Response function (log color scale) versus normalized frequency $\omega/\omega_J$ and linearly varied time delay  $\omega_JT$. The eigenvalues $\Omega_\alpha$ are computed for $N = 8$ cavity modes and are shown as superimposed white dashed lines. The parameters are chosen to be: $Z_0/Z_J = 100$ and $C_c/C_J = 0.3$.}
	\label{eigenvalues}
\end{figure}

We now go on to demonstrate that in the high impedance regime the response function $f(\omega)$ of the field trapped between the transmon and the mirror Eq.~\eqref{f_omega}, reproduces the dynamics of an effective Hamiltonian of a single transition atom in a multimode cavity. This Hamiltonian is of the form:
\begin{align}
\label{Hn}
H = \omega_J\left(a^\+a +\ZP\right) &+ \sum_{n=1}^\infty n \omega_c\left(c_n^\+c_n + \ZP\right)  \nonumber \\ &+ \frac{\Omega}{2} \sum_{n=1}^\infty  \left(a^\+ + a\right)\left(c_n^\+ + c_n\right),
\end{align}
%
%
%where here the cavity mode frequencies are given by $\omega_n = n\omega_C$, $\omega_C = 2\pi/T$ where $T = 2L/v$ is the time for a photon to travel twice the atom-mirror separation $L$, and the coupling strengths $g_n$ are given by the relation:
%\begin{align}
%	g_n = \frac{\Omega_n}{2},
%\end{align}
%with $\omega_B$ given in the text by Eq.~\eqref{eq:omegaB}. 
where the operators $a,a^\+$ are annihilation and creation bosonic operators ($[a,a^\+] = 1$) associated with excitation in the transmon qubit, while $c_n,c_n^\+$ annihilate/create photons in the cavity modes. When weakly excited, the choice for bosonic excitations of the transmon is justified, while the orthogonality relations between the cavity modes is ensured by the high finesse of the latter, so that we have $[c_n,c_m^\+] = \delta_{nm}$. Details of the diagonalisation of the Hamiltonian are shown in the supplemental materials. The response function $|f(\omega)|$ and eigenfrequencies of Hamiltonian \eqref{Hn} are shown in Fig.~\ref{eigenvalues}. The eigenfrequencies are shown to match the peaks of $|f(\omega)|$ for all cavity modes. Noticeably, a dip in the response function corresponding to the dark state is found for $\omega = \omega_0$.
\section{Discussion and Outlook}
We have made a first theoretical investigation of the properties of a transmon capacitively coupled to a high impedance transmission line, a system which is currently becoming experimentally accessible. By linearizing the Josephson junction, we could describe the low excitation dynamics, including spontaneous emission. We find qualitatively different behaviour, compared to the low impedance regime. In particular, the atom now forms its own cavity, and we can observe a vacuum Rabi splitting, giving rise to Rabi-oscillations in the spontaneous emission. The system is well described by a Hamiltonian for an atom weakly coupled to a multimode cavity.  We hope that this analysis will inspire an experimental realization of this novel system.  

\section{Acknowledgements}
The authors acknowledge funding from the Swedish Research Council (VR) through Grant No. 2016-06059. GJ also acknowledge funding from the Knut and Alice Wallenberg foundation (KAW) through the Wallenberg Centre for Quantum Technology (WACQT).

\bibliographystyle{apsrev4-2}
\bibliography{HighImpedance.bbl}

\newpage
\onecolumngrid
\appendix

%\preprint{APS/123-QED}

\section*{Supporting information for Transmon in a semi-infinite high-impedance transmission line \textemdash\ appearance of cavity modes and Rabi oscillations}
\begin{center}
E. Wiegand$^1$, B. Rousseaux$^2$ and G. Johansson$^1$\\
\emph{$^1$Department of Microtechnology and Nanoscience - MC2, Chalmers University of Technology, 412 96 G\"oteborg, Sweden}\\
\emph{$^2$Laboratoire de Physique de l'\'Ecole Normale Sup\'erieure, ENS, Universit\'e PSL, CNRS, Sorbonne Universit\'e, Universit\'e de Paris, F-75005 Paris, France}
\end{center}

\author{Emely Wiegand}

\affiliation{Department of Microtechnology and Nanoscience - MC2, Chalmers University of Technology, 412 96 G\"oteborg, Sweden}

\author{Benjamin Rousseaux}
\affiliation{Laboratoire de Physique de l'\'Ecole Normale Sup\'erieure, ENS, Universit\'e PSL, CNRS, Sorbonne Universit\'e, Universit\'e de Paris, F-75005 Paris, France}

\author{G\"oran Johansson}
\affiliation{Department of Microtechnology and Nanoscience - MC2, Chalmers University of Technology, 412 96 G\"oteborg, Sweden}

\maketitle
\vspace{.5cm}
%We provide supporting information about ... 

\section{Rabi oscillations of an atom coupled to a high-impedance semi-infinite TL - additional parameter values}
\begin{figure}[t]
	\flushleft
	\begin{minipage}{0.49\linewidth}
		\begin{overpic}[width=1\linewidth]{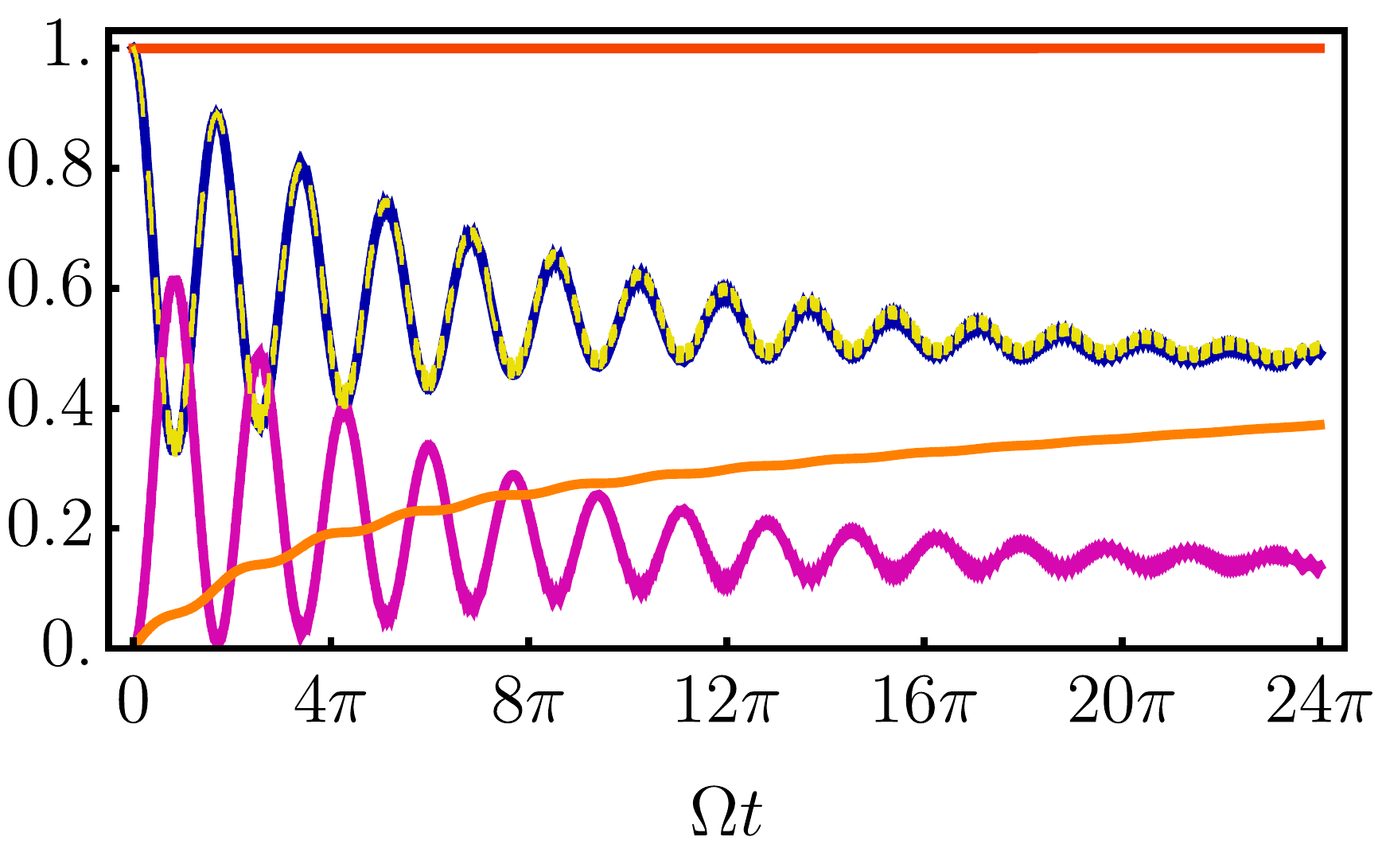}
			\put(-4,60){ $a)$}
			\put(98,60){ $b)$}
			\put(-4,-8){ $c)$}
			\put(98,-8){ $d)$}
		\end{overpic}
	\end{minipage}
	\begin{minipage}{0.49\linewidth}
		\includegraphics[width=1\linewidth]{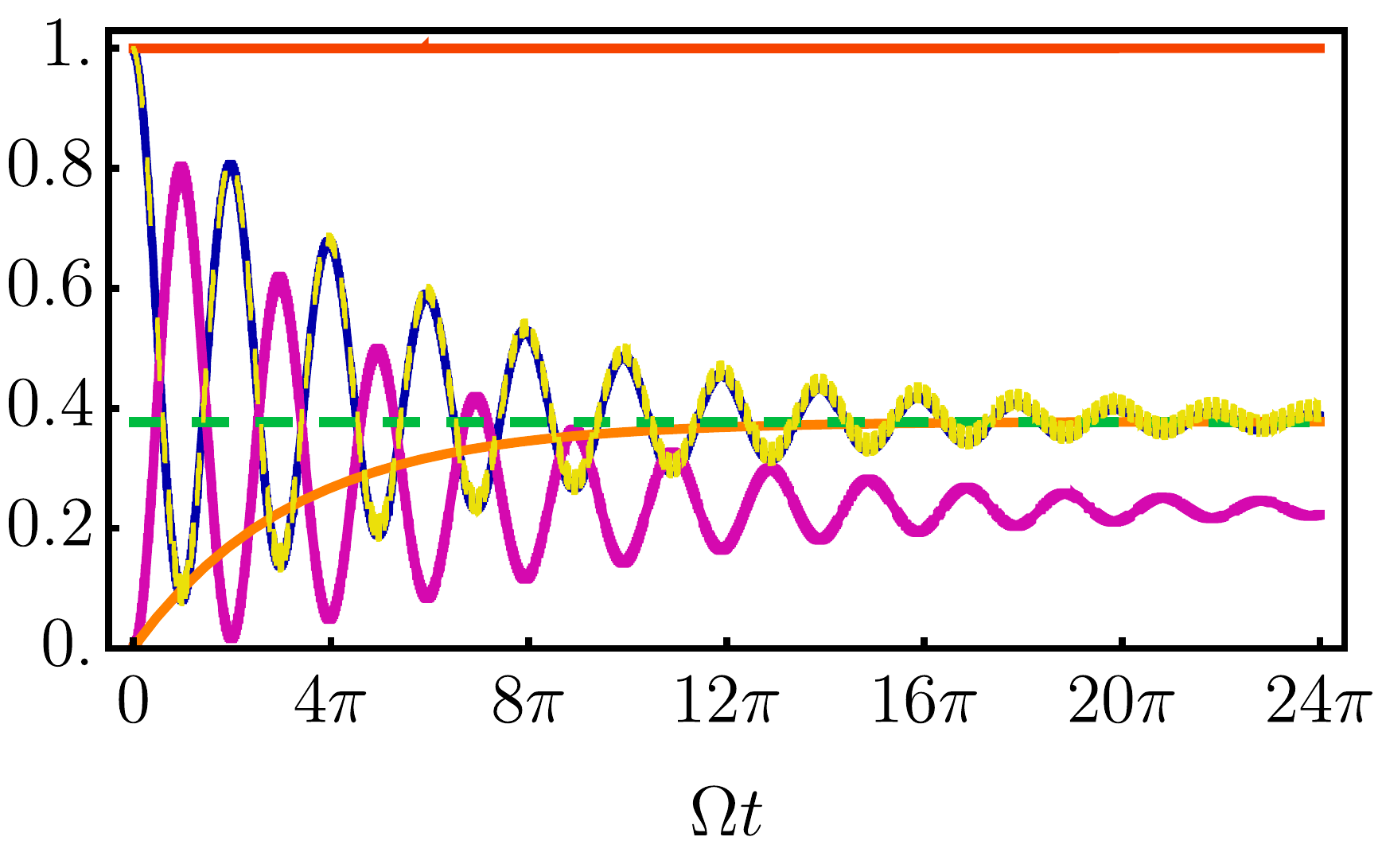}
	\end{minipage}
\flushleft
\begin{minipage}{0.49\linewidth}
	\includegraphics[width=1\linewidth]{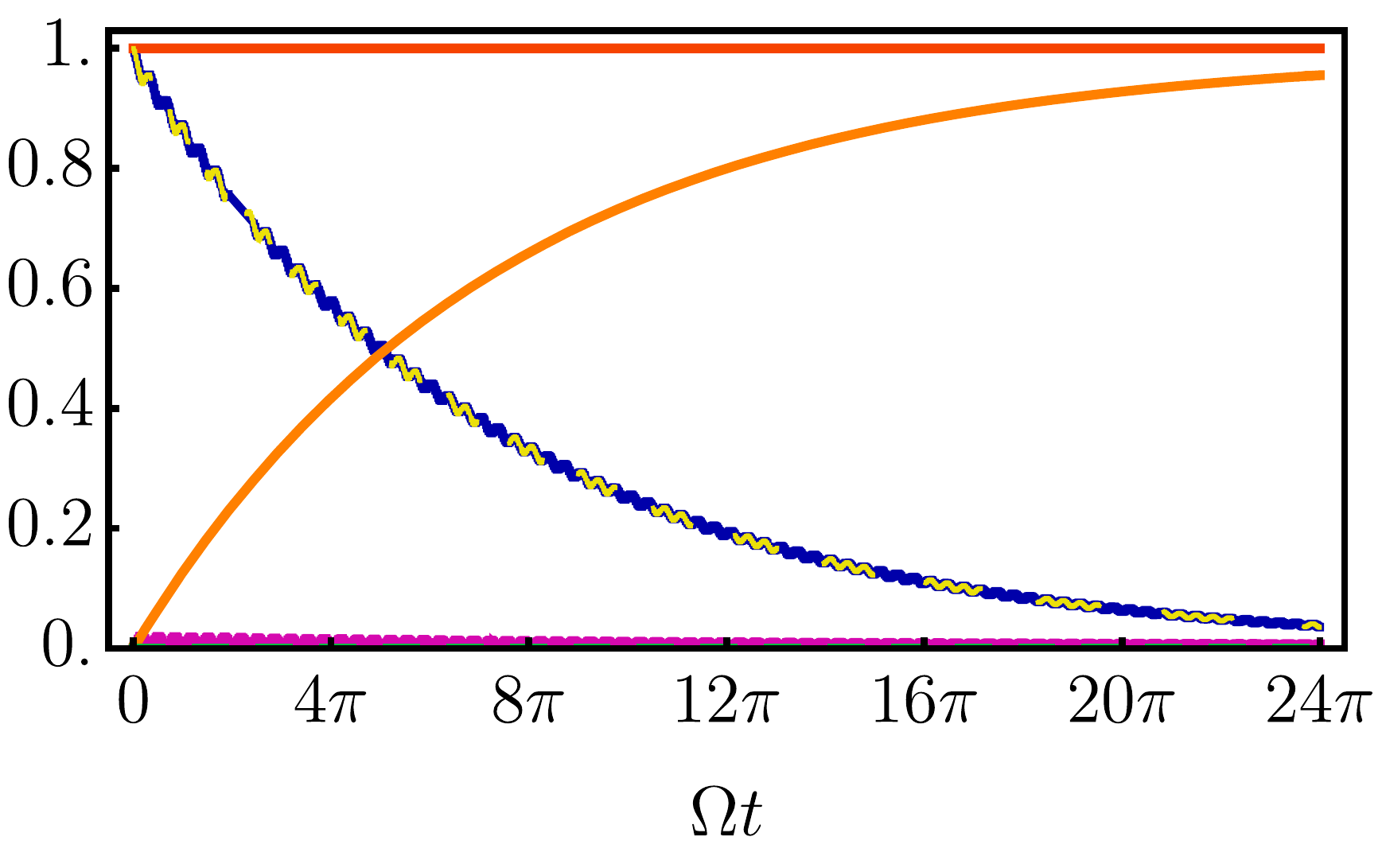}
\end{minipage}
\begin{minipage}{0.49\linewidth}
	\includegraphics[width=1\linewidth]{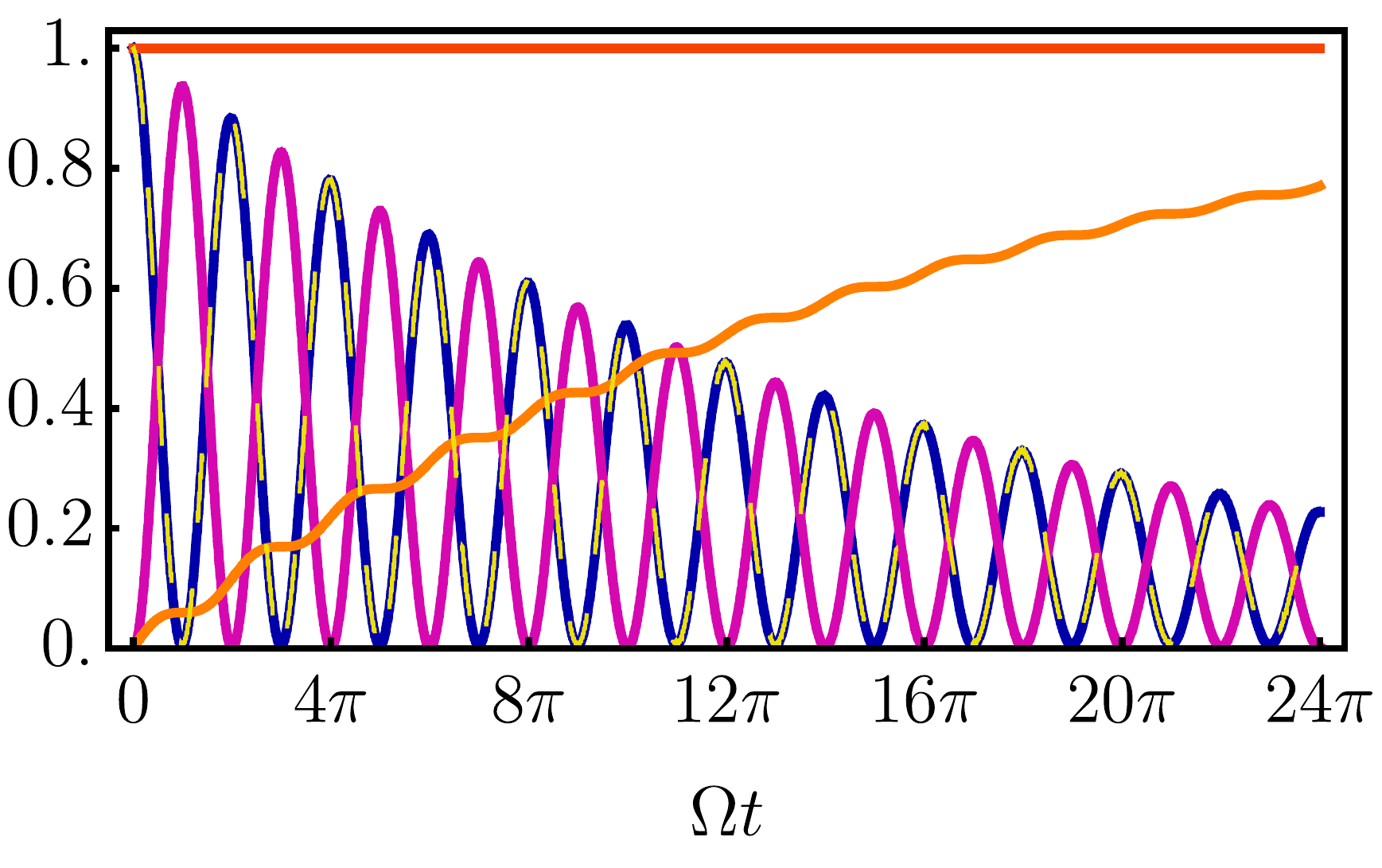}
\end{minipage}
\caption{Energy of the transmon (blue), energy of the field trapped between the transmon and the mirror (magenta), outging field to the left side of the transmon (orange), total energy of the system (red) and semi-numerically calculated energy of the transmon using Eq.~\eqref{eq:Laplace} as a function of period of the Rabi oscillations. The parameters for the panels are the following:
	a) $Z_0/Z_J = 1000$, $C_c/C_J = 0.02$ and $\omega_J = \omega_C$. Since we chose $C_c/C_J = 0.02$, the decay is very slow.
b) $Z_0/Z_J = 1000$, $C_c/(C_J+C_c) = 0.02$ and $\omega_0 = \omega_C$. Here, the parameters are almost the same as in a), but now the dark state condition is fulfiled, $\omega_J = \omega_0$. The system converges into a dark state with a finite excitation probability of both the transmon and the field between the transmon and the mirror. The green line indicates the analytical value of the dark state energy $\frac{E_{DS}}{E_0} =  \frac{1}{(1+ \frac{T }{2}\gamma_0)^2}$ with $\gamma_0 = \frac{Z_0 \omega_0^2}{2}\frac{C_c^2}{C_c + C_J}$.
c) $Z_0/Z_J = 1000$, $C_c/(C_J+C_c) = 0.3$ and $\omega_0 = \omega_C$. Here despite $\omega_0 = \omega_C$ as in b), the system does not seem to converge into a dark state and the decay rate is given by $\gamma_J = 1/C_J Z_0 $. The difference to b) is that the ratio between $\frac{C_c}{C_J} \frac{Z_0}{Z_J}$ here is much bigger than in b) which also means that $\omega_J$ is not close to $\omega_C$ and the Rabi oscillations are barely visible..
d) $Z_0/Z_J = 1000$, $C_c/C_J = 0.3$ and $\omega_J = \omega_C$. Here, as in c)  $\frac{C_c}{C_J} \frac{Z_0}{Z_J} \gg 1$  but the "Rabi condition" $\omega_J = \omega_C$ is still fulfilled. We see clear Rabi oscillations and the decay is slower compared to c) and the decay rate is given by $\gamma_J^m/2 = 1/2C_J Z_0 $.}
\label{fig:Rabi}
\end{figure}

\begin{figure}[t]
	\centering
	\begin{overpic}[width=0.7\linewidth]{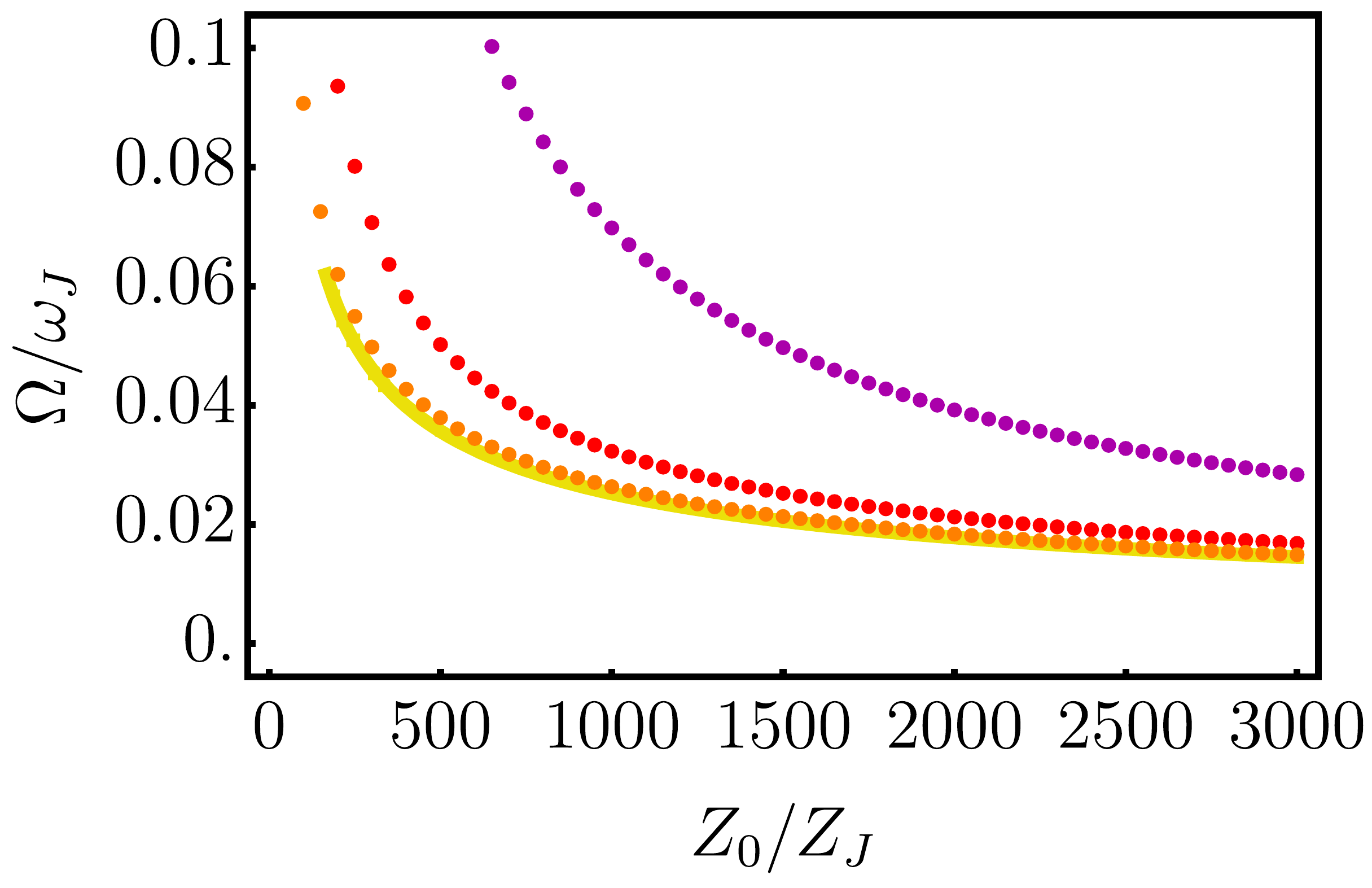} %HighImpedance.nb
		\put(68,58){$\Omega^N (\frac{C_c}{C_J} = 0.01)$}
		\put(68,52){$\Omega^N (\frac{C_c}{C_J} = 0.05)$}
		\put(68,45){$\Omega^N (\frac{C_c}{C_J} = 0.3)$}
		\put(68,39){$\Omega^A$}
	\end{overpic}
	\caption{Frequency of the beating $\Omega$ as a function of $Z_0/Z_J$ calculated analytically (yellow, solid) and numerically (dotted) for different values of $C_c/C_J$, $C_c/C_J=0.01$ (purple), $C_c/C_J=0.05$ (red), $C_c/C_J=0.3$ (orange). We see that the higher $\frac{C_c}{C_J} \frac{Z_0}{Z_J}$ becomes, the closer the approximated analytical frequency $\Omega^A$ and the numerically calculated frequency $\Omega^N$ become.
	}
	\label{fig:OmegaB}
\end{figure}

 We can Laplace transform the equations of motion of the transmon shown in the main article and extract the exact poles numerically. With the following formulas we can calculate the inverse Laplace transform of the system variables and energy:
\begin{align}
\text{Res}_{1,2}^{\pm} p_J(s) = \lim_{s \rightarrow s_{1,2}^{\pm}} p_J(s) (s - s_{1,2}^{\pm}),
\end{align}
\begin{align}
    p_J(t) = 	\int p_J(s) e^{-s t} d s =  \left.2 \pi i \sum_{k} \operatorname{Res} p_J(s) e^{-s t} \right|_{s=s_{k}}
    \label{eq:Laplace}
\end{align}
where $k = s_{1,2}^{\pm}$ are the poles of $p_J(s)$. Similarly, we calculate $\phi_J(t)$. We show the results as an addition to Fig.~4 in the main article. Here we provide more figures for different system parameters.
In all panels of Fig~\ref{fig:Rabi}, the impedance is chosen to be $Z_0/Z_J = 1000$. In Fig.~\ref{fig:Rabi} a), the ratio of the coupling capacitance and the Josephson capacitance is fairly small $C_c/C_J = 0.02$ and the coupling to the TL is weak. The cavity frequency equals the resonance frequency of the transmon $\omega_C= \omega_J$. The decay is weak and the Rabi-oscillations are clearly visible.  The parameters in Fig.~\ref{fig:Rabi} b) are similar to a), but now $C_c/(C_J+C_c) = 0.02$ and most importantly %the atom is located at a node of the electromagnetic field, which means that 
the cavity frequency equals the transition frequency of the qubit for low impedance, $\omega_C= \omega_0$, which is the condition for a dark state \cite{Wiegand2020}. The energy of the transmon decays until it reaches the dark state, with energy $\frac{E_{DS}}{E_0} =  \frac{1}{(1+ \frac{T }{2}\gamma_0)^2}$ where $\gamma_0 = \frac{Z_0 \omega_0^2}{2}\frac{C_c^2}{C_c + C_J}$ \cite{Wiegand2020}. In c), the coupling capacitance is much larger compared to a) and b), $C_c/(C_c + C_J) = 0.3$ and the transmon fulfills the dark state condition
%is located at a node 
$\omega_C = \omega_0$. Anyhow, the system does not converge into a dark state and the Rabi oscillations are very weak. The main difference here is that, since $C_c/(C_c + C_J)$ is rather large, it means that $\omega_J$ is not close to $\omega_C$ and the Rabi oscillations and coupling to the cavity are suppressed. In this parameter regime, the behaviour of the system seems to be independent of the position of the transmon with respect to the mirror. Similar to Fig.~\ref{fig:Rabi} c), in Fig.~\ref{fig:Rabi} d), the coupling capacitance is rather large too $C_c/C_J = 0.3$, but here the cavity frequency equals the frequency of the transmon $\omega_C = \omega_J$, which means that the Rabi condition is fulfilled. We see clear Rabi oscillations and in addition the decay is much slower than in $c)$.
As mentioned in the main article, we find an analytical expression for the oscillation frequency in the high impedance regime by analyzing the Laplace transform of the equations of motion. We find the Rabi frequency to be $\Omega = \frac{  2}{ \sqrt{T C_J Z_0} }$. In Fig.~\ref{fig:OmegaB}, we demonstrate the deviations of the approximation from numerically calculated values. We find that the higher the ratio $\frac{C_c}{C_J} \frac{Z_0}{Z_J}$, the closer the approximation resembles the numerical solution.
\section{Analysis of the response functions}
To analyze the solution of the Fourier transformation of the equations of motion in the main article in a convenient manner, we introduce the following functions
\begin{align}
	N (\omega) &= R_0 (\omega) - i R_J (\omega), \\
	R_0 (\omega) &= \left( 1 - \frac{\omega^2}{\omega_0^2} \right) \\
	R_J (\omega) &= \frac{C_c  Z_0}{2} \omega \left( 1 - \frac{\omega^2}{\omega_J^2} \right).
\end{align}
With these definitions we are able to write the transmission and the reflection amplitudes of the qubit in the open transmission line as 
%\textcolor{red}{Sign?}
%
%
%
\begin{align}
	r (\omega) = \frac{i R_J (\omega)}{ N (\omega)}, \quad
	t(\omega) = \frac{R_0 (\omega)}{N (\omega)}, \quad r(\omega) + 1 = t (\omega).
\end{align}
We note that the high impedance regime corresponds to that $|R_J(\omega)| \gg |R_0(\omega)|$ away from resonances, i.e. $|C_c Z_0 \omega /2 | \gg 1$, while the opposite ($|R_J(\omega)| \ll |R_0(\omega)|$) is true in the low impedance regime $|C_c Z_0 \omega /2 | \ll 1$. 
\subsection{Damping rate for the open TL}
We analyse the scattering solution of the qubit excitation $\phi_J$ to find the damping rate for the transmon in an open TL. The solution in frequency space reads
\begin{align}
	\phi_J(\omega) = - \frac{C_c L_J \omega}{R_0 (\omega) - R_J (\omega)}.
\end{align}
In the high impedance regime, we perform an expansion around the bare qubit frequency $\omega = \omega_J + \delta \omega$ and find to first order in $\delta \omega$
\begin{align}
	\phi_J (\omega_J+\delta \omega) \approx \frac{i}{\omega_J}      \frac{1} {1 - i 2 \delta \omega/\gamma_J},
\end{align}
where $\gamma_J = 2/Z_0 C_J$ is the energy damping rate for spontaneous emission. 

In the low impedance regime, the qubit resonance is shifted to $\omega_0$ and we instead expand $\omega = \omega_0 + \delta \omega$ to find
\begin{align}
	\phi_J (\omega_0+\delta \omega)  = \frac{1}{C_c Z_0 \omega_0^2} \frac{1}{1 - i 2 \delta \omega/ \gamma_0},
\end{align}
where $\gamma_0 = \frac{Z_0 \omega_0^2}{2} \frac{C_c^2}{C_c + C_J} $ is the low-impedance damping rate \cite{Wiegand2020}. %\textcolor{red}{Actually is half of it. cite our old paper}.
\subsection{Damping rates and Lamb shifts with a mirror}

With a mirror, the solution for the Josephson flux $\phi_J$
%, the charge on the JJ $p_J$ and the response function $f(\omega)$ 
can be written as
\begin{align}
		\phi_J (\omega)  &=  \frac{  -i C_c L_J \omega \left(  1 - e^{i \omega T} \right) }{ R_0 (\omega) - i R_J (\omega)  \left(  1 - e^{ i \omega T} \right)  }, %\\
%		p_J (\omega) &= \frac{   C_c  \omega^2 \left(  e^{i \omega T}  - 1 \right) }{ R_0 (\omega) - i R_J (\omega)  \left(  1 - e^{ i \omega T} \right)  }, \\
%		f(\omega) &= \frac{1}{1 - i \frac{R_J (\omega)}{R_0 (\omega) }\left( 1 - e^{i \omega T}  \right)}.
\end{align}
%
%
%
%
%
%
%
%
%
%
%
%\subsection{Damping rate with the mirror}%
%
%
%
In the low impedance regime, we find that the qubit resonance is Lamb-shifted to $\tilde \omega_0 = \omega_0 + \gamma_0 \sin(\omega_0 T)/2$ and we can expand $\phi_J(\omega)$ around the resonance and find
\begin{align}
	\phi_J (\tilde{\omega}_0 +\delta \omega) \approx -i \frac{C_c}{C_J+C_c} \frac{e^{i\omega_0 T/2}}{\gamma_0 \sin \left( \omega_0 T/2\right)} \frac{1}{1 - i 2\delta \omega / \gamma_0^m},
\end{align}
%
%
%where $\gamma_0^m = 2 \gamma_0 \left( 1 - e^{i \omega_0 T}  \right)$ is the damping rate of the transmon in front of a mirror. The real and imaginary parts give rise to the Lamb Shift and Purcell effect, respectively.
%For small $Z_0$, we find
%
%
%\begin{align}
%	\phi_J (\tilde \omega_0 + \delta \omega) = \frac{C_c L_J \omega_0^2 \sin \left( \omega_0 T /2\right)  e^{i \omega_0 T/2}  }{i 2 \gamma_0 \sin^2 \left( \omega_0 T/2 \right)}  \frac{1}{1 - i \frac{\omega - \omega_0}{\tilde \gamma_0^m}},
%\end{align}
%
%
with the Purcell-modified damping rate $\gamma_0^m = 2 \gamma_0 \sin^2 \left( \omega_0 T/2\right)$. 

In the high impedance regime, the resonance frequency is Lamb shifted to $\tilde{\omega}_J=\omega_J+\gamma_J \cot{\left(\omega_J T /2\right)}/4$. Expanding $\phi_J(\omega)$ around this frequency we find
%
%
%
%\begin{align}
%	\phi_J (\tilde{\omega}_J + \delta \omega) = - \frac{1}{\omega_J} \frac{1}{ C_c Z_0 \delta \omega - e^{i \omega_0 T/2}  \sin^{-1} \left( \omega_0 T /2  \right)}.
%\end{align}
%
%
%The resonance, which corresponds to the Lamb shift is given at \textcolor{red}{Cite something with Lambshift}
%
%
%\begin{align}
%	\delta \omega = \frac{1}{2 C_J Z_0} \cot \left(  \frac{\omega_J T}{2} \right).
%\end{align}
%
%Now we can evolve $\phi_J$ around this value and find
%
%
%
\begin{align}
	\phi_J (\tilde{\omega}_J + \delta \omega) = - \frac{2 i}{ \omega_J} \frac{1}{1 + i 2 \delta \omega /\gamma_J^m},
\end{align}
where $\gamma_J^m = \gamma_J/2=1/C_J Z_0$ is the damping rate of an atom in front of a mirror in the high $Z_0$ regime. Here, we note that the expression for the Lamb-shift diverges when $\sin{\left(\omega_J T /2\right)}=0$, i.e. when $\omega_J$ is close to a cavity resonance $\omega_C^n = 2 \pi n / T$. This is when the single pole approximation is no longer valid and we find the vacuum Rabi splitting. Away from the Rabi condition, we also note that the damping rate is independent of the distance to the mirror, i.e. we see no Purcell effect. 
Away from the Rabi condition, we can also analyze the response function  
\begin{align}
%		\phi_J (\omega)  &=  \frac{  -i C_c L_J \omega \left(  1 - e^{i \omega T} \right) }{ R_0 (\omega) - i R_J (\omega)  \left(  1 - e^{ i \omega T} \right)  }, \\
%		p_J (\omega) &= \frac{   C_c  \omega^2 \left(  e^{i \omega T}  - 1 \right) }{ R_0 (\omega) - i R_J (\omega)  \left(  1 - e^{ i \omega T} \right)  }, \\
		f(\omega) &= \frac{1}{1 - i \frac{R_J (\omega)}{R_0 (\omega) }\left( 1 - e^{i \omega T}  \right)},
\end{align}
to extract the cavity modes.
Due to the finite transmission through the transmon, they are slightly shifted from the perfect mirror frequencies to $\tilde{\omega}_c^n=\omega_c^n + R_0(\omega_c^n)/T R_J(\omega_c^n)$. Close to the resonances we can expand $\omega=\tilde{\omega}_c^n+\delta \omega$ and find
\begin{align}
	f (\tilde{\omega}_c^n+\delta \omega) \approx \frac{2}{t (\omega_C^n)} \frac{1}{ 1 - i2\delta \omega / \gamma_c^n},
\end{align}
where $\gamma_c^n \approx |t(\tilde \omega_C^n)|^2/T$ is the energy damping rate of cavity mode $n$.
%  \section{Effective quantum model: atom in a multimode cavity Hamiltonian}
%
%
%

%
%
%
%
%\textcolor{red}{Shorten the text here since it is in the main article already}  It is shown that the response function of the trapped field in between the transmon and the mirror (Eq. (7) in the main text), reproduces the dynamics of an effective Hamiltonian resembling that of a single transition atom in a multimode cavity in the limit of high ratio $Z_0/Z_J$. This Hamiltonian is of the form:
%  \begin{align}
%  H = \omega_J\left(a^\+a +\ZP\right) + \sum_{n=1}^\infty\omega_n\left(c_n^\+c_n + \ZP\right) + \sum_{n=1}^\infty g_n\left(a^\+ + a\right)\left(c_n^\+ + c_n\right),
%  \end{align}
%where here the cavity mode frequencies are given by $\omega_n = n\omega_C$, $\omega_C = 2\pi/T$ where $T = 2L/v$ is the time for a photon to travel twice the atom-mirror separation $L$, and the coupling strengths $g_n$ are given by the relation:
%\begin{align}
%g_n = \frac{\omega_B}{\sqrt{n}},
%\end{align}
%with $\omega_B$ given in the text by Eq. (13). The operators $a,a^\+$ are annihilation and creation bosonic operators ($[a,a^\+] = 1$) associated with excitation in the transmon qubit, while $c_n,c_n^\+$ annihilate/create photons in the cavity modes formed by the mirror and the transmon. When weakly excited, the choice for bosonic excitations of the transmon is justified, while the orthogonality relations between the cavity modes is ensured by the high finesse of the latter, so that we have $[c_n,c_m^\+] = \delta_{nm}$. The Hamiltonian can be numerically diagonalized with the following procedure.

\section{Hopfield diagonalization of the atom in a mulimode cavity Hamiltonian}
To diagonalize the effective quantum optical Hamiltonian (Eq.~(12) in the main article), we first introduce the new polaritonic operators:
\begin{align}
	\Pi_\alpha = x^\alpha a + y^\alpha a^\+ +\sum_n\left(m_n^\alpha c_n + h_n^\alpha c_n^\+\right),
\end{align}
$x^\alpha,y^\alpha,m_n^\alpha,h_n^\alpha$ being the Hopfield coefficients associated to each bosonic operator $a,a^\+,c_n,c_n^\+$. To ensure the bosonicity of the polaritonic operators ($[\Pi_\alpha,\Pi_\beta^\+] = \delta_{\alpha\beta}$), the coefficients should satisfy the relation:
\begin{align}
	|x^\alpha|^2 - |y^\alpha|^2 + \sum_n\left(|m_n^\alpha|^2 - |h_n^\alpha|^2\right) = 1.
\end{align}
The new operators should satisfy the eigenvalue problem:
\begin{align}
	\left[\Pi_\alpha,H\right] = \Omega_\alpha\Pi_\alpha,
\end{align}
where $\Omega_\alpha$ are the eigenfrequencies labeled with a new index $\alpha$. Typically, if we couple the atom with $N$ cavity modes, then $\alpha$ runs from 1 to $N+1$. Expanding the commutator in the previous equation, it is possible to write the eigenvalue problem in a matrix form:
\begin{align}
	\left(
	\begin{array}{ccccccc}
	\omega_J & 0 & g_1 & -g_1 & \hdots & g_N & -g_N\\
	0 & -\omega_J & g_1 & -g_1 & \hdots & g_N & -g_N\\
	g_1 & -g_1 & \omega_1 & 0 & \hdots & 0 & 0\\
	g_1 & -g_1 & 0 & -\omega_1 &  & 0 & 0\\
	\vdots & \vdots & \vdots &  & \ddots & & \vdots \\
	g_N & -g_N & 0 & 0 &  & \omega_N & 0\\
	g_N & -g_N & 0 & 0 & \hdots & 0 & -\omega_N
	\end{array}\right)\left(\begin{array}{c}x^\alpha\\ y^\alpha\\m_1^\alpha\\h_1^\alpha\\\vdots\\m_N^\alpha\\h_N^\alpha\end{array}\right) = \Omega_\alpha\left(\begin{array}{c}x^\alpha\\ y^\alpha\\m_1^\alpha\\h_1^\alpha\\\vdots\\m_N^\alpha\\h_N^\alpha\end{array}\right).
\end{align}
This eigenvalue problem can be solved analytically for $N=1$ but in general, one has to diagonalize it numerically. A comparison between the full response function and the eigenvalues are shown in Fig.~5 in the main article.
% \ref{eigenvalues}.
%\begin{figure}
%    \centering
%    \includegraphics[scale = .7]{f_omega_v5.pdf}
%    \caption{Response function (log color scale) versus normalized frequency $\omega/\omega_J$ and linearly varied cavity frequency $\omega_C/\omega_J$. The eigenvalues $\Omega_\alpha$ are computes for $N = 5$ cavity modes and are shown as superimposed white dashed lines. The parameters are chosen to be: $Z_0/Z_J = 100$ and $C_c/C_J = 0.5$.}
%    \label{eigenvalues}
%\end{figure}

\end{document}